\documentclass{aa} 

\usepackage{amsmath,amssymb,latexsym,mathrsfs}
\usepackage{epsfig}
\usepackage{natbib}
\usepackage{txfonts}
\bibpunct{(}{)}{;}{a}{}{,} 
\usepackage{graphicx}
\usepackage{longtable}
\usepackage{booktabs}
\usepackage{float}
\usepackage{dcolumn}
\usepackage{graphics,epsfig}
\usepackage[breaklinks, colorlinks, citecolor=blue]{hyperref}
\usepackage{bm}
\usepackage{color}

\begin{document}

\title{A WISE view of the composite luminosity function of cluster members}


\author{C. De Boni
       \inst{1,2}
       \and
       N. Aghanim\inst{2}
       \and
       M. Douspis\inst{2}
       \and
       E. Soubri\'e\inst{2}
       \and
       R. De Propris \inst{3}
       }

\institute{Max-Planck-Institut f\"ur extraterrestrische Physik (MPE), Gie{\ss}enbachstra{\ss}e, D-85741 Garching bei M\"unchen, Germany
\and
Institut d'Astrophysique Spatiale, CNRS (UMR8617) Universit\'e Paris-Sud 11, B\^{a}t. 121, F-91405 Orsay, France
\and
FINCA, University of Turku, V{\"a}is{\"a}l{\"a}ntie 20, Piikki{\"o}, FI-21500, Finland \\
\email{cdeboni@mpe.mpg.de}
}

   \date{Received MM DD, YYYY; accepted MM DD, YYYY}

 
\abstract
{Galaxy clusters play a crucial role in constraining cosmological parameters. Cluster detection and characterization have therefore become an important field of modern cosmology.}
{In this work, we aim at constructing a reasonably representative composite luminosity function of cluster members in the WISE survey that can be used in the future as an input in matching algorithms for cluster detection.}
{We use a sample of massive clusters from the redMaPPer SDSS catalog to match the positions of the WISE sources. We build the composite luminosity function of the WISE members in different redshift bins.}
{We find that galaxy cluster members have a characteristic composite luminosity function, with a clear change in the slope at a given apparent magnitude $M^{*}$, which becomes fainter with increasing redshift. The best-fit bright-end slope $\beta$ is compatible with a constant value of $3.54$ with no clear trend with redshift, while the faint-end slope $\alpha$ is remarkably different, ranging from $2$ to $3$.} 
{We present the first characterization of the composite luminosity function of the WISE counterparts of redMaPPer SDSS cluster members and in this way we provide an element for building cluster detection techniques based on matching algorithms.}


\maketitle
%

\section{Introduction}

Galaxy clusters are fundamental tools in modern cosmology. Their distribution in number and mass as a function of redshift is an indicator of the background cosmological properties of the Universe [{\it e.g.} \cite{2016A&A...594A..24P,2017arXiv170800697S} and references therein]. For this reason the detection of statistically complete samples of these objects is of considerable importance for their use as cosmological probes.

Galaxy clusters are complex objects containing a hot intracluster gas (ICM), amounting to about 15\% of their mass, and galaxies, amounting to about 5\%, lying in a dark matter (DM) potential well which represents about 80\% of the total mass. Clusters emit at several wavelengths and so are detectable by different means. X-ray catalogs [{\it e.g.} REFLEX \citep{2004A&A...425..367B}] are constructed using the X-ray emission from the hot plasma of the ICM. The same ICM can be detected through the Sunyaev-Zel'dovich (SZ) effect \citep{1972CoASP...4..173S}. Recently, the Planck \citep{2016A&A...594A..27P,2011A&A...536A...8P}, ACT \citep{2013JCAP...07..008H} and SPT \citep{2015ApJS..216...27B} produced several catalogs of SZ clusters. In total, they contain more than $1700$ identified SZ-selected clusters up to $z=1.5$ (\url{http://szcluster-db.ias.u-psud.fr/}). 
Optical catalogs [{\it e.g.} redMaPPer SDSS \citep{2014ApJ...785..104R, 2016ApJS..224....1R}] use the information from the emission of the stellar component to detect clusters. There are several ways to identify clusters in optical observations. redMaPPer is a red-sequence cluster finder optimized for large photometric surveys like SDSS (see \cite{2015ApJS..219...12A} for recent releases). For another red-sequence based detection algorithm, see \cite{2017arXiv170504331L}. There exist also cluster detection techniques based on matching filter algorithms for photometric optical/infrared (IR) data [{\it e.g.} BCF \citep{2012MNRAS.420.1167A} and AMICO \citep{2017arXiv170503029B}]. In general, those algorithms try to match a template for the object under detection to the distribution of sources. For clusters, the template is usually a combination of the spatial distribution {\it e.g.}, Gaussian, NFW \cite{1996ApJ...462..563N} or $\beta$-model \cite{1976A&A....49..137C} profile and the luminosity function of the galaxy members.

The luminosity function (LF) is a common tool to characterize the distribution of luminous matter in the Universe [{\it e.g.} \cite{2003MNRAS.342..725D,2016MNRAS.459.3998L,2017AJ....153..189L} and references therein]. It depends on the environment, and its evolution with redshift can be used to trace the changes in the galaxy population. The exact shape of the LF is still debated, in particular regarding the faint end [{\it e.g.} \cite{2005ApJ...633..122H,2009A&A...508.1217Z,2011MNRAS.414.2771D,2015A&A...581A..11M} and references therein]. There is general agreement that the bright end is described by a Schechter function \citep{1976ApJ...203..297S}. At optical and infrared wavelengths, it is usually determined from photometry by statistical means, either in a single band or using color information. In particular, the infrared $K$ band provides a good approximation of the underlying stellar mass function \citep{1996A&A...312..397G,2001ApJ...550..212B,2018MNRAS.473..776K}. Because the bright end is always poorly sampled in a single cluster, individual LF of different objects can be combined in a composite luminosity function \citep[][and references therein]{2003MNRAS.342..725D}.

In this paper, we aim at characterizing the IR luminosity function of cluster members using the Wide-field Infrared Survey Explorer (WISE) data \citep{2010AJ....140.1868W}. WISE is an infrared space telescope which surveyed the all-sky in four mid-infrared bands at $3.4$, $4.6$, $12$ and $22 \ \mu m$. By completely scanning the sky twice, it took images of three-quarter of a billion objects, ranging from galaxies to stars and asteroids (\url{https://www.nasa.gov/mission_pages/WISE/mission/index.html}). We stress here that our main purpose is to provide a description of how cluster galaxies appear in WISE, so that the cluster galaxy LF we obtain may be used in the future in matched-filter detection techniques applied to WISE data. Numerous other studies of galaxy clusters focused on their IR properties using the WISE data to detect new clusters. The MaDCoWS project \citep[][and references therein]{2014ApJS..213...25S} aims at detecting high-redshift massive galaxy clusters by combining WISE and SDSS data and using spectroscopic observations to confirm their results. \cite{2017AstL...43..507B} extended the Planck cluster catalog by identifying clusters combining the $3.4 \ \mu m$ band of WISE and the SDSS data, using the algorithm described in \cite{2015AstL...41..167B}. For rich ($\lambda > 40$, where $\lambda$ is the richness parameter defined in Section \ref{data}) redMaPPer clusters, the algorithm detects IR emission in WISE images convolved with $\beta$-models and then identifies the brightest central galaxy and the red sequence using SDSS data. In both cases, no information on the luminosity function is used. Indeed, the cluster members luminosity function has remained mostly unexplored in WISE despite its potential importance as an ingredient in cluster detection techniques based on matched filtering. By using WISE data, we are facing several issues (e.g. lack of redshifts; difficulty in identifying cluster members and the Brightest Cluster Galaxy; contamination by foreground galaxies, AGNs, etc.). Therefore, our approach is to choose a  large sample of known clusters with identified members from the redMaPPer catalog and to characterize their IR properties in WISE through the LF. This provides us with a reasonably representative LF to be used in the future cluster detections. WISE and SDSS use different magnitude systems (VEGA and AB, respectively), but this is not going to affect our results because we only use SDSS objects to crossmatch the WISE catalog.

The paper is organized as follows: in Section \ref{data} we describe the data, the sample selection and the matching procedure. The construction of the WISE composite luminosity function and the parametrization we use to fit it are presented in Section \ref{WISE_LF}. In Section \ref{results} we describe the our results. We discuss their impact and draw our conclusions in Section \ref{conclusions}. Throughout the paper, we use the following cosmological parameters: $H_{0}=70 \ {\rm km s^{-1} Mpc^{-1}}$, $\Omega_{m0}=0.3$ and $\Omega_{\Lambda}=0.7$.
   

\section{Data} \label{data}

The first step to build the luminosity function is to identify cluster members in WISE. 

\subsection{Cluster samples}

We start from the redMaPPer DR8 cluster catalog of the SDSS survey (\url{http://risa.stanford.edu/redmapper/}). It contains $26311$ groups and clusters with redshifts ranging from $0.08$ to $0.6$ detected using a photometric red-sequence cluster finder algorithm \citep{2016ApJS..224....1R}. Every galaxy is associated a redMaPPer photometric redshift probability $p_{mem}$ \citep{2014ApJ...785..104R}. The cluster richness $\lambda$ is defined as the sum of the membership probabilities over all galaxies within a scale-radius. The spectroscopic probability $p_{spec}$ is defined by comparing the membership probability with the measured spectroscopic membership rate \citep{2015MNRAS.453...38R}. The only other quantities we are taking from the SDSS catalog, in addition to the above mentioned ones, are the celestial coordinates of cluster members. 

\begin{figure}
\centering
\hbox{
 \includegraphics[width=0.45\textwidth]{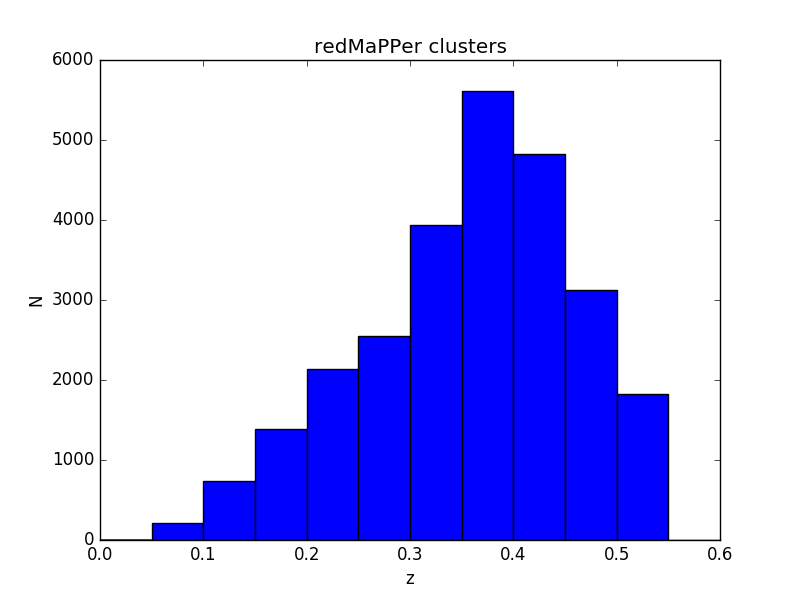}
}
\hbox{
}
\caption{Redshift distribution of redMaPPer SDSS clusters.}
\label{rozo_redshift}
\end{figure}

Given the redshift distribution of the redMaPPer clusters (see Figure \ref{rozo_redshift}), we consider nine redshift bins of width $\Delta z=0.05$, from $z=0.10-0.15$ up to $z=0.50-0.55$. We define a reference cluster sample constituted by $100$ randomly selected massive clusters in each redshift bin by setting the richness limit to $\lambda > 45$.

\begin{figure}
\centering
\hbox{
 \includegraphics[width=0.45\textwidth]{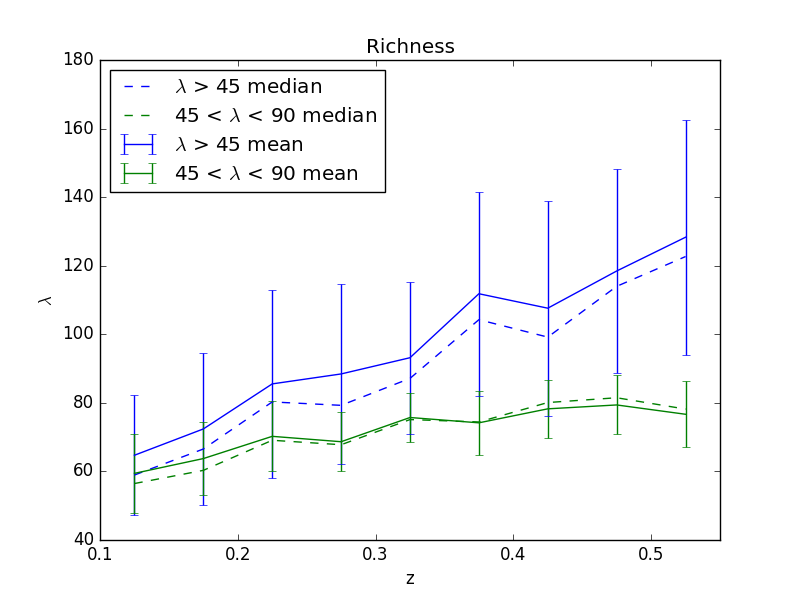}
}
\hbox{
}
\caption{Mean (full line) and median (dashed line) richness as a function of redshift for the $\lambda > 45$ (blue) and $45 < \lambda < 90$ (green) samples. Error bars mark one standard deviation.}
\label{richness}
\end{figure}

We consider also a second sample from which we excluded the most massive clusters by randomly choosing $100$ objects with $45 < \lambda < 90$, roughly corresponding to a mass range from $2.6$ to $6.4 \times 10^{14} h^{-1} {\rm M_{\odot}}$ \citep{2017MNRAS.466.3103S,2017arXiv170701907M}. This control sample will allow us to check that our results are not driven by the most massive clusters. In both cluster samples, the average richness increases with redshift, but as expected the effect is less pronounced in sample 2, when we do not consider very rich systems (see Fig. \ref{richness}).

\subsection{Galaxy members}

For each cluster in the two samples, we search for the WISE counterparts of the galaxy members. The redMaPPer catalog provides us with the galaxy members in SDSS. For every member in the redMaPPer catalog with spectroscopic probability $p_{spec} > 0.5$ we compute the celestial coordinate distance and search the closest counterpart in the WISE galaxy catalog. Using $p_{spec}$ instead of $p_{mem}$ is a more conservative choice, because there are few members with $p_{mem}>0.5$ and $p_{spec}<0.5$, but not the reverse, so by choosing $p_{spec}>0.5$ we also ensure that $p_{mem}>0.5$. We checked that the cross-match completeness did not change from $p_{spec}>0.9$ to $p_{spec}>0.5$, so we choose this value in order to increase the statistics of the galaxy members in each cluster. We retain as matched members only WISE sources that have a distance to the SDSS source $< 0.5 \ {\rm arcsec}$. The point spread function (PSF) of WISE is $6$", while the PSF of SDSS is $1.4$". This can generate blending of different SDSS sources into a single WISE source. For this reason, we allow multiple assoctiations of different SDSS sources to a single WISE source, but not the contrary. We stress however that the frequency of these multiple matches is rare, and this should not affect our results. For the spatially matched WISE sources, we consider the magnitude measured with profile-fitting photometry, noted $mpro$  (see \url{http://wise2.ipac.caltech.edu/docs/release/allsky/expsup/sec2_2a.html#w1mpro}). For our study, we are mainly interested in the magnitudes in the $w1$ band of WISE, corresponding to $3.4 \ \mu m$, since it is the most sensitive to star emission from galaxies. All the same, we consider in addition the $w2$ band at $4.6 \ \mu m$, which is also sensitive to hot dust. We further set a last constraint to the cross-match between SDSS cluster members and WISE sources. The errors on both $w1$ and $w2$ magnitudes should be greater than $0$. This ensures us that we have a measured magnitude in both bands and not only an upper limit. We do not perform any color selection, but we check {\it a posteriori} that the contamination by quasar objects is low by verifying that sources with $w1 - w2 > 0.4$ represent only a few percent of the matched members \citep[cfr.][]{2016A&A...592A..25K}.

\begin{figure}
\centering
\hbox{
 \includegraphics[width=0.45\textwidth]{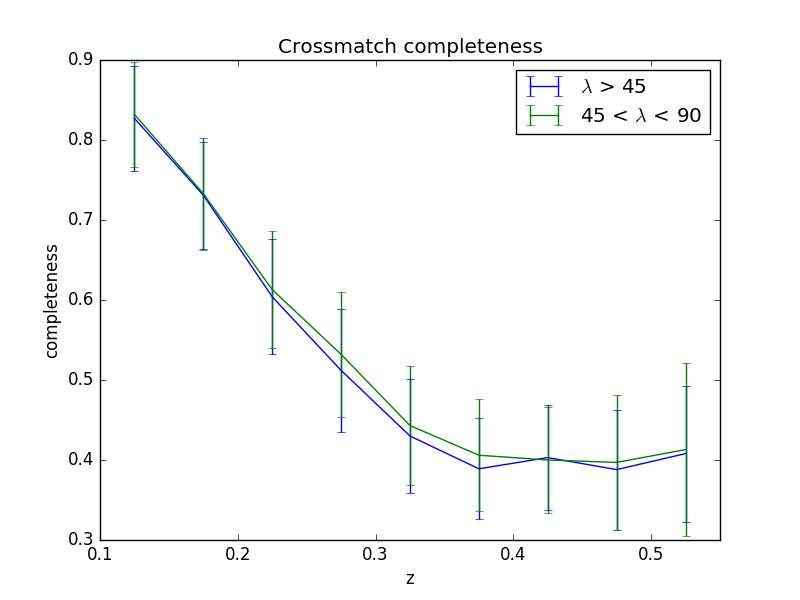}
}
\hbox{
}
\caption{Mean crossmatch completeness as a function of redshift for the $\lambda > 45$ (blue) and $45 < \lambda < 90$ (green) samples. Error bars mark one standard deviation.}
\label{completeness}
\end{figure}

For each cluster from the two samples, we can define a completeness as the ratio between the number of retained matched WISE sources and the initial number of redMaPPer SDSS members (with $p_{spec} > 0.5$). While it is important to bear in mind that this ratio can depend on luminosity, we however describe the behaviour of the obtained ``effective" completeness. We find that it is a decreasing function of redshift independent of the richness of the sample. Indeed, as can be seen from Fig. \ref{completeness}, the evolution of the effective completeness with redshift is identical for the two cluster samples: it decreases linearly from approximately $80 \%$ at $z=0.10$ to $40 \%$ at $z=0.35$, and then remains constant. So at $z>0.3$ we lose more than half of the initial redMaPPer members, but this fraction is independent of the richness of the sample, which is significantly different at those redshifts.

\section{The WISE composite luminosity function}\label{WISE_LF}

Following \cite{2003MNRAS.342..725D}, we construct the apparent $w1$ and $w2$ composite luminosity function  \citep{1989MNRAS.237..799C} for the WISE counterparts of SDSS cluster members. The number of galaxies in the $j$-th magnitude bin of the composite LF, $N_{cj}$, is defined as:

\begin{equation}
N_{cj}=\frac{N_{c0}}{m_j}\sum_{i}\frac{N_{ij}}{N_{i0}} \ ,
\label{Ncj}
\end{equation}

\noindent where $N_{ij}$ is the number of galaxies in the $j$-th bin of the $i$-th cluster LF, $N_{i0}$ is the normalization used for the $i$-th cluster LF (number of galaxies brighter than $w1_{norm}$), $m_j$ is the number of clusters contributing to the $j$-th bin, and $N_{c0}$ is the sum of all the normalizations. The latter is given by:

\begin{equation}
N_{c0}=\sum_{i}N_{i0} \ .
\end{equation}

\noindent The errors in $N_{cj}$ are computed according to

\begin{equation}
\delta N_{cj} = \frac{N_{c0}}{m_j} \left[ \sum_{i} \left( \frac{\delta N_{ij}}{N_{i0}} \right)^2  \right]^{1/2} \ ,
\end{equation}

\noindent where $\delta N_{cj}$ and $\delta N_{ij}$ are the errors in the $j$-th LF bin for the composite and $i$-th cluster, respectively. In our case, we assign Poissonian errors to $N_{ij}$, so that $\delta N_{ij} = \sqrt{N_{ij}}$ and

\begin{equation}
\delta N_{cj} = \frac{N_{c0}}{m_j} \left( \sum_{i} \frac{N_{ij}}{N_{i0}^2}  \right)^{1/2} \ .
\label{sigma}
\end{equation}

As discussed in the introduction, by using WISE data alone we do not have any information on the redshifts of the sources. For this reason, and with the aim of detecting clusters without having any a priori knowledge of their redshift, we consider here apparent magnitudes. We will briefly discuss absolute magnitudes in Appendix \ref{app_absolute}. For the same reason, we cannot correct the effective completeness for redshift effects as in \cite{2003MNRAS.342..725D}. In order to compute the number of galaxies in the $j$-th magnitude bin of the composite LF, we consider a magnitude bin width of $0.5 \ {\rm mag}$. This is large enough to contain several galaxies within each bin and small enough to sample sufficiently well the LF by allowing a sufficient number of bins. For the normalization, we set $w1_{\mathrm{norm}} = 17$ and $w2_{\mathrm{norm}} = 16$. Those values are close to the limiting magnitude of WISE, but the precise choice has not a crucial impact on our final results. For example, the shape of the bright end composite LF for $w1$ does not change for $w1_{\mathrm{norm}}$ in the range $[16-18]$.

Our objective is not to explore in detail the physical properties of the WISE LF of the redMaPPer SDSS clusters but rather to characterize the composite LF, in particular its shape, in order to apply it as a template for blind detection of clusters in WISE. We therefore fit the normalized composite LF, i.e. the composite LF divided by the number of clusters used to build it, $N_{cj,norm}$. This is similar, in principle, to what is done in \cite{2015A&A...581A..11M} and leads to a LF which is representative of the average cluster.
The WISE composite luminosity function increases with increasing magnitudes, up to a point when it starts decreasing. This feature was already evident in the SDSS composite luminosity function. It reflects the incompleteness of the original data and it is not due to the crossmatching procedure. In Figure \ref{composite_LF_w1} we show the obtained composite LFs (points with error bars) of the reference sample for the $w1$ band, up to the point where they reach their maximum before decreasing.

We now fit the obtained composite LF. Different fitting functions have been proposed in the literature for the IR luminosity function [{\it e.g.} \cite{2005ApJ...632..169L,2006A&A...445...29P,2006MNRAS.370.1159B,2017arXiv170207829L} and references therein]. We found that the Schechter function was not returning a good fit to our data. For this reason, in our study, we consider a double power law \citep{2006MNRAS.370.1159B,2010A&A...514A...6G} which is known to provide a good description of the cluster members IR luminosity function: 

\begin{eqnarray}
\phi (L) dL &=& \left( \frac{\phi^{*}}{L^{*}} \right) \left( \frac{L}{L^{*}} \right)^{1-\alpha} dL \ \ \ \ \ \ \ (L<L^{*}) \nonumber \\ 
\phi (L) dL &=& \left( \frac{\phi^{*}}{L^{*}} \right) \left( \frac{L}{L^{*}} \right)^{1-\beta} dL \ \ \ \ \ \ \ (L>L^{*}) \end{eqnarray}

\noindent leading to

\begin{eqnarray} 
\phi (M) dM &=& 0.4 \ \ln 10 \ \phi^{*} \ 10^{0.4 (2-\alpha) (M^{*}-M)} dM \ \ \ \ \ \ \ (M>M^{*}) \nonumber \\
\phi (M) dM &=& 0.4 \ \ln 10 \ \phi^{*} \ 10^{0.4 (2-\beta) (M^{*}-M)} dM \ \ \ \ \ \ \ (M<M^{*}) \ ,
\label{double_power_law_m}
\end{eqnarray}

\noindent where $\alpha$ is the slope of the faint end and $\beta$ is the slope of the bright end. The other two parameters of the fit are the normalization $\phi^{*}$ and the transition magnitude between the two power laws $M^{*}$.

In each redshift bin, we fit the composite LF up to the magnitude where it reaches its maximum, and do not consider data at fainter magnitudes. We evaluate the quality of the fit by computing the reduced $\chi^2_{\mathrm{red}}$ of the fit as

\begin{equation}
\chi^2_{\mathrm{red}} = \frac{1}{N_{\mathrm{bin}}-N_{\mathrm{par}}} \sum_{N_{\mathrm{bin}}} {\frac{({\mathrm{data}}-{\mathrm{fit}})^2}{\sigma^2}} \ ,
\end{equation}

\noindent where $\sigma$ is given by equation (\ref{sigma}) and $N_{par}=4$.

We noted that the fit is sensitive to the choice of the initial parameters, so we sample the parameter space with an iterative process, starting from a default set of initial values (namely, $\phi^*_{ini} = max(N_{cj,norm})$, $M^*_{ini} = w(\phi^*_{ini})$, $\alpha_{ini} = 2$ and $\beta_{ini} = 2$). For the first iteration, we use the best-fit parameters as starting values for the new fit. Then, we compare the $\chi^2_{\mathrm{red}}$ of the two fits and choose the parameters giving the best $\chi^2_{\mathrm{red}}$ as initial parameters for the following iteration. We perform $10^4$ iterations and take the final results as best-fit parameters. We note that in general the fit converges after a number of iterations which is orders of magnitude smaller than the total number of iteration. For every redshift bin, we independently repeat the iteration $10$ times to avoid any bias in the fit.


\section{Results} \label{results}

\begin{figure}
\centering
\hbox{
 \includegraphics[width=0.45\textwidth]{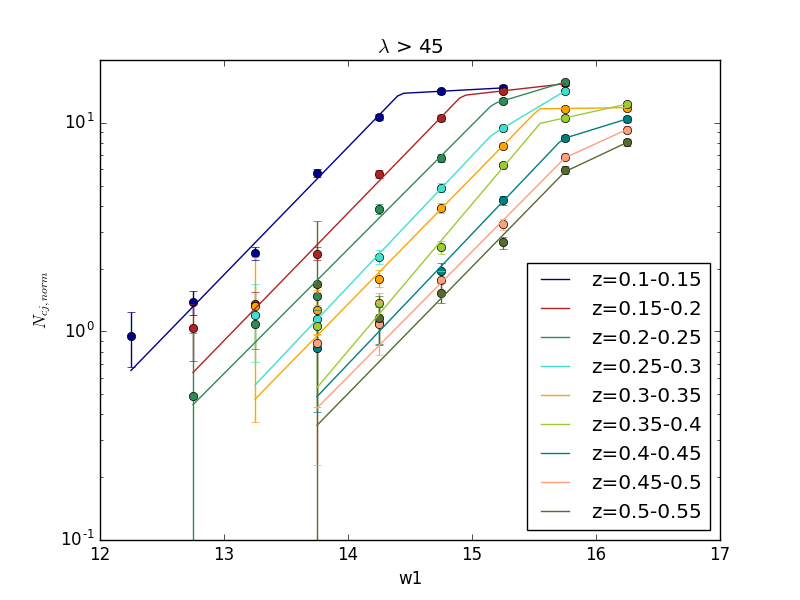}
}
\hbox{
}
\caption{Normalized composite luminosity function (points with error bars) $N_{cj,norm}$, as defined in Section \ref{WISE_LF}, in the nine redshift bins, for our reference sample, $\lambda > 45$, in the $w1$ band. Full lines show the best-fit of the double power law formula, equation (\ref{double_power_law_m}), to the data.}
\label{composite_LF_w1}
\end{figure}

Figure \ref{composite_LF_w1} displays the obtained composite LFs for $w1$ magnitudes of our reference sample. It is immediately evident that there is a break in the slope at a given magnitude. For each redshift bin, we provide in Tables \ref{w1_17_45-500}, \ref{w1_17_45-90}, \ref{w2_16_45-500} and \ref{w2_16_45-90} the best-fit parameters and associated errors for the $w1$ and $w2$ composite luminosity function of the two cluster samples. In the following, we will not discuss the normalization $\phi^{*}$ but rather focus on the shape parameters of the composite LF.

\begin{table*}
\begin{center}
\caption{Best-fit parameters for the double power law, equation (\ref{double_power_law_m}), fit to the normalized apparent $w1$ composite LF for WISE data (with $w1_{norm} = 17$) for our reference sample ($\lambda>45$). \label{w1_17_45-500}}
\begin{tabular}{cccccccccc}
\hline
\hline
$z$ & $\phi^{*}$ & $\sigma_{\phi^{*}}$ & $\alpha$ & $\sigma_{\alpha}$ & $\beta$ & $\sigma_{\beta}$ & $M^{*}$ & $\sigma_{M^{*}}$ & $\chi^2_{red}$ \\
\hline
$0.10-0.15$ & $15.01$ & $0.77$ & $2.08$ & $0.08$ & $3.53$ & $0.06$ & $14.42$ & $0.05$ & $2.26$ \\
\hline
$0.15-0.20$ & $14.67$ & $0.82$ & $2.17$ & $0.08$ & $3.53$ & $0.06$ & $14.92$ & $0.05$ & $2.22$ \\ 
\hline
$0.20-0.25$ & $13.39$ & $0.76$ & $2.45$ & $0.09$ & $3.49$ & $0.09$ & $15.17$ & $0.08$ & $2.36$ \\ 
\hline
$0.25-0.30$ & $9.51$ & $1.40$ & $2.89$ & $0.10$ & $3.57$ & $0.14$ & $15.16$ & $0.15$ & $0.95$ \\ 
\hline
$0.30-0.35$ & $12.67$ & $0.57$ & $2.02$ & $0.09$ & $3.52$ & $0.09$ & $15.54$ & $0.05$ & $0.83$ \\ 
\hline
$0.35-0.40$ & $10.79$ & $0.64$ & $2.33$ & $0.09$ & $3.76$ & $0.12$ & $15.55$ & $0.06$ & $1.54$ \\ 
\hline
$0.40-0.45$ & $9.09$ & $0.64$ & $2.46$ & $0.11$ & $3.56$ & $0.17$ & $15.73$ & $0.11$ & $0.62$ \\ 
\hline
$0.45-0.50$ & $7.38$ & $--$ & $2.68$ & $--$ & $3.50$ & $0.10$ & $15.75$ & $--$ & $0.84$ \\ 
\hline
$0.50-0.55$ & $6.33$ & $--$ & $2.70$ & $--$ & $3.52$ & $0.11$ & $15.75$ & $--$ & $2.47$ \\ 
\hline
\hline
\end{tabular}
\end{center}
\end{table*}

\begin{table*}
\begin{center}
\caption{Best-fit parameters for the double power law, equation (\ref{double_power_law_m}), fit to the normalized apparent $w1$ composite LF for WISE data (with $w1_{norm} = 17$) for our control sample ($45<\lambda<90$). \label{w1_17_45-90}}
\begin{tabular}{cccccccccc}
\hline
\hline
$z$ & $\phi^{*}$ & $\sigma_{\phi^{*}}$ & $\alpha$ & $\sigma_{\alpha}$ & $\beta$ & $\sigma_{\beta}$ & $M^{*}$ & $\sigma_{M^{*}}$ & $\chi^2_{red}$ \\
\hline
$0.10-0.15$ & $13.49$ & $0.76$ & $2.13$ & $0.09$ & $3.50$ & $0.06$ & $14.41$ & $0.05$ & $1.67$ \\
\hline
$0.15-0.20$ & $13.13$ & $0.74$ & $2.18$ & $0.09$ & $3.46$ & $0.06$ & $14.95$ & $0.05$ & $1.50$ \\
\hline
$0.20-0.25$ & $11.40$ & $0.70$ & $2.45$ & $0.09$ & $3.36$ & $0.10$ & $15.22$ & $0.10$ & $2.12$ \\
\hline
$0.25-0.30$ & $6.78$ & $1.06$ & $2.88$ & $0.11$ & $3.60$ & $0.15$ & $15.02$ & $0.14$ & $1.96$ \\
\hline
$0.30-0.35$ & $10.31$ & $0.53$ & $2.02$ & $0.10$ & $3.52$ & $0.10$ & $15.52$ & $0.06$ & $0.67$ \\
\hline
$0.35-0.40$ & $7.71$ & $0.46$ & $2.17$ & $0.12$ & $3.39$ & $0.13$ & $15.59$ & $0.08$ & $0.36$ \\
\hline
$0.40-0.45$ & $6.68$ & $--$ & $2.25$ & $--$ & $3.44$ & $0.09$ & $15.75$ & $--$ & $1.57$\\ 
\hline
$0.45-0.50$ & $5.15$ & $--$ & $2.43$ & $--$ & $3.25$ & $0.10$ & $15.75$ & $--$ & $0.26$ \\ 
\hline
$0.50-0.55$ & $3.90$ & $--$ & $2.63$ & $--$ & $3.17$ & $0.14$ & $15.75$ & $--$ & $2.13$ \\ 
\hline
\hline
\end{tabular}
\end{center}
\end{table*}

\begin{table*}
\begin{center}
\caption{Best-fit parameters for the double power law, equation (\ref{double_power_law_m}), fit to the normalized apparent $w2$ composite LF for WISE data (with $w2_{norm} = 16$) for our reference sample ($\lambda>45$). \label{w2_16_45-500}}
\begin{tabular}{cccccccccc}
\hline
\hline
$z$ & $\phi^{*}$ & $\sigma_{\phi^{*}}$ & $\alpha$ & $\sigma_{\alpha}$ & $\beta$ & $\sigma_{\beta}$ & $M^{*}$ & $\sigma_{M^{*}}$ & $\chi^2_{red}$ \\
\hline
$0.10-0.15$ & $12.79$ & $0.64$ & $2.26$ & $0.09$ & $3.65$ & $0.09$ & $14.07$ & $0.06$ & $0.78$ \\
\hline
$0.15-0.20$ & $11.48$ & $0.82$ & $2.40$ & $0.09$ & $3.61$ & $0.09$ & $14.43$ & $0.06$ & $1.48$ \\ 
\hline
$0.20-0.25$ & $14.74$ & $0.68$ & $2.04$ & $0.09$ & $3.45$ & $0.07$ & $15.00$ & $0.05$ & $0.09$ \\ 
\hline
$0.25-0.30$ & $12.28$ & $0.58$ & $2.16$ & $0.09$ & $3.58$ & $0.09$ & $15.08$ & $0.06$ & $0.86$ \\ 
\hline
$0.30-0.35$ & $9.59$ & $0.67$ & $2.44$ & $0.10$ & $3.46$ & $0.13$ & $15.14$ & $0.10$ & $1.15$ \\ 
\hline
$0.35-0.40$ & $8.25$ & $0.87$ & $2.65$ & $0.11$ & $3.58$ & $0.18$ & $15.23$ & $0.14$ & $1.05$ \\ 
\hline
$0.40-0.45$ & $9.30$ & $0.47$ & $2.11$ & $0.11$ & $3.49$ & $0.12$ & $15.62$ & $0.07$ & $0.25$ \\ 
\hline
$0.45-0.50$ & $7.78$ & $0.44$ & $2.25$ & $0.12$ & $3.40$ & $0.17$ & $15.73$ & $0.12$ & $1.03$ \\
\hline
$0.50-0.55$ & $6.34$ & $--$ & $2.63$ & $--$ & $3.34$ & $0.10$ & $15.75$ & $--$ & $1.45$ \\ 
\hline
\hline
\end{tabular}
\end{center}
\end{table*}

\begin{table*}
\begin{center}
\caption{Best-fit parameters for the double power law, equation (\ref{double_power_law_m}), fit to the normalized apparent $w1$ composite LF for WISE data (with $w2_{norm} = 16$) for our control sample ($\lambda>45$). \label{w2_16_45-90}}
\begin{tabular}{cccccccccc}
\hline
\hline
$z$ & $\phi^{*}$ & $\sigma_{\phi^{*}}$ & $\alpha$ & $\sigma_{\alpha}$ & $\beta$ & $\sigma_{\beta}$ & $M^{*}$ & $\sigma_{M^{*}}$ & $\chi^2_{red}$ \\
\hline
$0.10-0.15$ & $11.57$ & $0.62$ & $2.28$ & $0.09$ & $3.60$ & $0.10$ & $14.06$ & $0.06$ & $0.68$ \\
\hline
$0.15-0.20$ & $10.22$ & $0.77$ & $2.42$ & $0.09$ & $3.53$ & $0.09$ & $14.44$ & $0.07$ & $0.86$ \\
\hline
$0.20-0.25$ & $12.13$ & $0.61$ & $2.03$ & $0.09$ & $3.37$ & $0.07$ & $14.98$ & $0.05$ & $1.19$ \\
\hline
$0.25-0.30$ & $9.49$ & $0.54$ & $2.19$ & $0.10$ & $3.46$ & $0.10$ & $15.06$ & $0.07$ & $0.61$ \\
\hline
$0.30-0.35$ & $8.24$ & $0.52$ & $2.33$ & $0.11$ & $3.42$ & $0.14$ & $15.17$ & $0.10$ & $1.31$ \\
\hline
$0.35-0.40$ & $5.96$ & $--$ & $2.51$ & $--$ & $3.27$ & $0.09$ & $15.25$ & $--$ & $0.40$ \\
\hline
$0.40-0.45$ & $6.53$ & $0.41$ & $2.09$ & $0.13$ & $3.30$ & $0.14$ & $15.60$ & $0.09$ & $0.82$ \\ 
\hline
$0.45-0.50$ & $5.41$ & $--$ & $2.14$ & $--$ & $3.12$ & $0.09$ & $15.75$ & $--$ & $0.32$ \\ 
\hline
$0.50-0.55$ & $4.13$ & $--$ & $2.42$ & $--$ & $3.01$ & $0.11$ & $15.75$ & $--$ & $0.98$ \\ 
\hline
\hline
\end{tabular}
\end{center}
\end{table*}

\begin{figure}
\centering
\hbox{
 \includegraphics[width=0.45\textwidth]{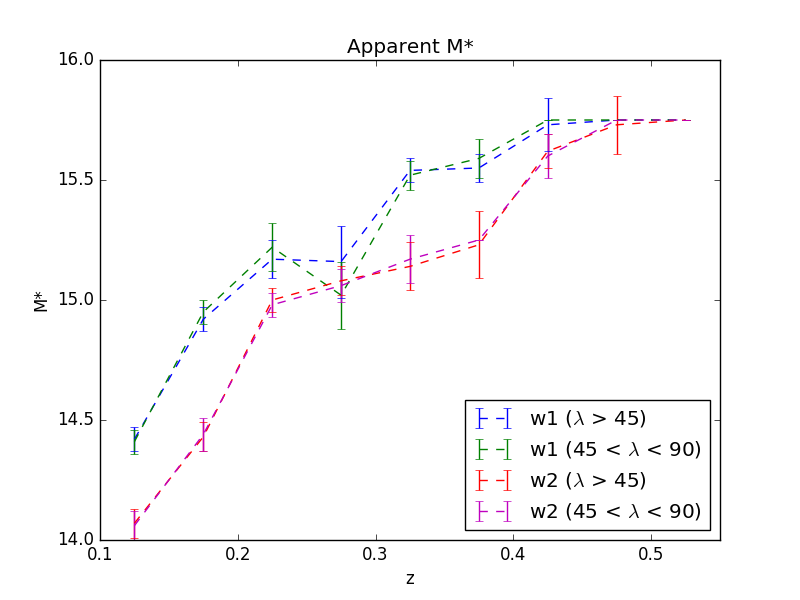}
}
\hbox{
}
\caption{Evolution with redshift of the best-fit apparent $M^{*}$ for $w1$ ($\lambda>45$) [blue dashed-line], $w1$ ($45<\lambda<90$) [green dashed-line], $w2$ ($\lambda>45$) [red dashed-line] and $w2$ ($45<\lambda<90$) [magenta dashed-line]. Error bars mark one standard deviation.}
\label{w_app_star}
\end{figure}

We first focus on the apparent magnitude $M^{*}$ marking the transition between the two regimes of the double power-law composite LF. For our reference cluster sample, we note that the best-fit value of $M^{*}$ for the $w2$ LF are 0.5 magnitudes brighter than the best-fit values for the $w1$ LF. We also note that the best-fit values of $M^{*}$ for the $w1$ LF are the same, within the error bars, when considering the reference sample ($\lambda > 45$) and the sample without the richest clusters ($45 < \lambda < 90$), and the same is true for $w2$. This indicates that $M^{*}$ is no biased by the most massive clusters. Finally as can be seen from Fig. \ref{w_app_star}, the best-fit $M^{*}$ increases with redshift, with high-redshift clusters showing the transition between the two LF regimes one magnitude fainter than low-redshift clusters. We note here, but discuss in Appendix \ref{app_absolute}, that this is not just a redshift effect.

\begin{figure*}
\centering
\hbox{
 \includegraphics[width=0.45\textwidth]{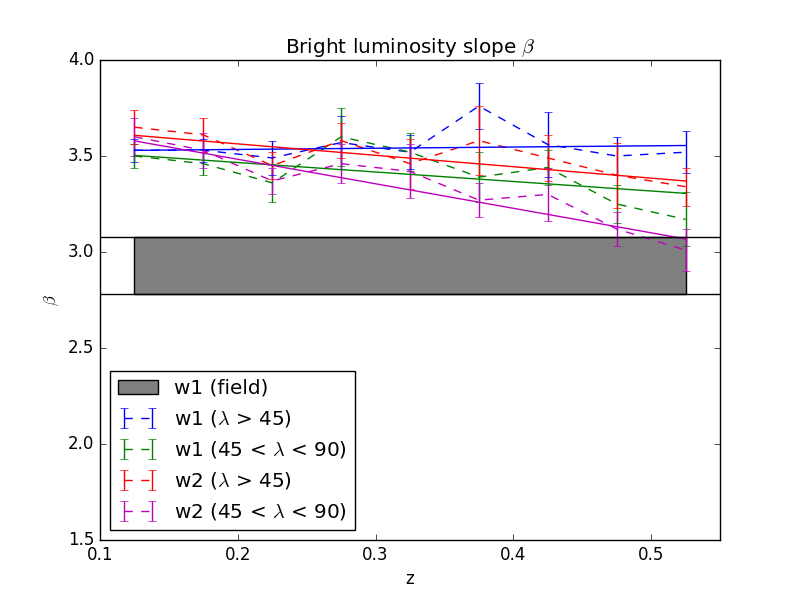}
 \includegraphics[width=0.45\textwidth]{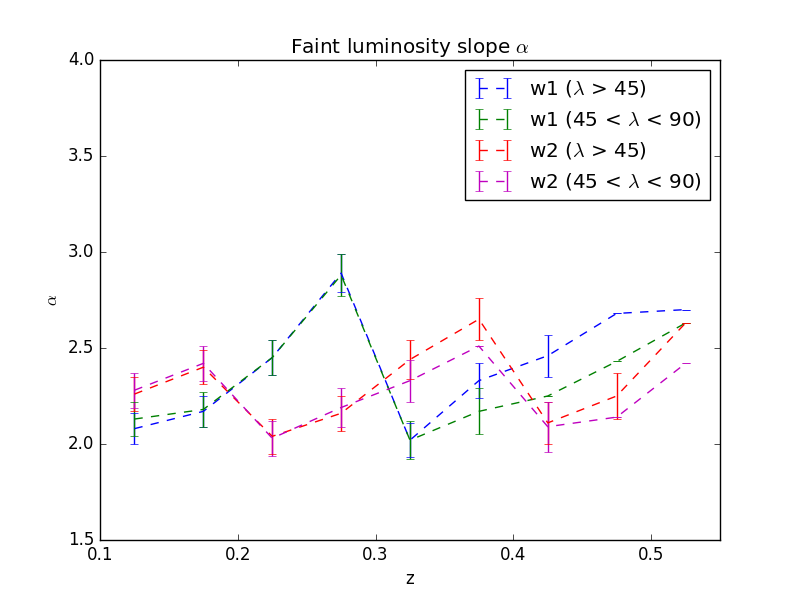}
}
\hbox{
}
\caption{Evolution with redshift of the best-fit bright $\beta$ (left panel) and faint $\alpha$ (right panel) end slopes for $w1$ ($\lambda>45$) [blue dashed-lines], $w1$ ($45<\lambda<90$) [green dashed-lines], $w2$ ($\lambda>45$) [red dashed-lines] and $w2$ ($45<\lambda<90$) [magenta dashed-line]. The black horizontal lines in the left panel mark the best-fit value of $\beta$ for a $1 \ {\rm deg}^2$ field centered on the same clusters and for a random field of the same size. Error bars mark one standard deviation. The full lines in the left panel show the best-fit of the linear evolution of $\beta$ with redshift in the three cases.}
\label{slopes}
\end{figure*}

We now turn to the slopes of the composite LF. The obtained best-fits values are summarized in Tables \ref{w1_17_45-500}, \ref{w1_17_45-90}, \ref{w2_16_45-500} and \ref{w2_16_45-90} and displayed in Fig. \ref{slopes}. The left panel shows the results for $\beta$, i.e. the slope of the power law in the bright end of the composite LF up to $M^{*}$. Let us start by considering the reference sample in the $w1$ band. The values for $\beta$ range between $3.49$ and $3.76$. If we fit the redshift evolution of $\beta$ with a constant [$\beta(z) = \beta_0$] we get a value of $3.54$, while if we fit it with a linear relation [$\beta(z) = \beta_0 + s z$] we get a result compatible with no evolution, and the two $\chi^2$ are similar, as summarized in Table \ref{beta_evolution}. So we can conclude that the bright end slope of the composite LF of our reference sample is constant with redshift. If we now move to the second sample, the one that does not contain the richest clusters, we see that, while the fit with a constant still provides an acceptable $\chi^2$, the linear relation provides a better fit, with a negative slope incompatible with no redshift evolution. A similar result is found when considering the two samples in the $w2$ band. In both samples, the evolution with redshift is stronger for $w2$ than for $w1$. Most importantly, the value of $\beta$ for the cluster composite LF significantly differs from the average value for the field. We checked this by evaluating the field composite LF for $1 \ {\rm deg}^2$ patches of the sky in the same regions where the clusters of the reference sample are located, and by repeating the procedure in other $10$ random patches of the sky of the same size, for a total of $910$ field patches. We do this simply by evaluating the composite LF of all the sources in the field satisfying the selection criteria we used for the cluster members. The slope of the power law for the field varies within $2.8$ and $3.1$, with no significant scatter and no evolution with redshift. This result clearly shows that the cluster LF of our reference sample shows a characteristic bright end in the WISE $w1$ band, steeper than the one of the field, with a constant slope of $3.5$. Even when considering the second sample of clusters, despite the fact that the redshift evolution of $\beta$ is better described by a decreasing linear function, the bright-end slope is incompatible with the one of the field.

\begin{table*}
\begin{center}
\caption{Best-fit parameters for the constant and linear function fit to the evolution of the bright-end slope $\beta$ with redshift. \label{beta_evolution}}
\begin{tabular}{cccccccccccc}
\hline
\hline
Band & $\lambda$ & & \multicolumn{3}{c}{Constant} & & \multicolumn{5}{c}{Linear} \\
\multicolumn{2}{c}{} & & $\beta_{0}$ & $\sigma_{\beta_{0}}$ & $\chi^2$ & & $\beta_{0}$ & $\sigma_{\beta_{0}}$ & $s$ & $\sigma_s$ & $\chi^2$ \\
\hline
$w1$ & $\lambda > 45$ & & $3.54$ & $0.03$ & $0.50$ & & $3.52$ & $0.07$ & $+0.06$ & $0.22$ & $0.56$ \\
\hline
$w1$ & $45 < \lambda < 90$ & & $3.43$ & $0.03$ & $1.37$ & & $3.56$ & $0.07$ & $-0.49$ & $0.22$ & $0.87$ \\
\hline
$w2$ & $\lambda > 45$ & & $3.51$ & $0.03$ & $1.08$ & & $3.68$ & $0.08$ & $-0.59$ & $0.27$ & $0.52$ \\
\hline
$w2$ & $45 < \lambda < 90$ & & $3.34$ & $0.03$ & $3.60$ & & $3.74$ & $0.08$ & $-1.28$ & $0.25$ & $0.47$ \\
\hline
\hline
\end{tabular}
\end{center}
\end{table*}

The right panel of Fig. \ref{slopes} shows the best-fit values for the faint-end slope, $\alpha$, of the composite LF. The values for $\alpha$ range from $2$ to $3$ and are significantly different from the ones for $\beta$, indicating that there is indeed a transition in the shape of the LF for cluster members, that justify the use of the double power law equation \ref{double_power_law_m}. This is not the case for the field, where the value for $\alpha$ is similar to $\beta$, around $3$, meaning that the field is well described by a single power law. Even if it is difficult to find a trend with redshift of $\alpha$ (but we note here that the value is similar for the two samples, indicating that the average richness has no major impact on these results, and is different for $w1$ and $w2$), it is the change itself in the slope that can be used as another indicator to identify cluster members in WISE. Indeed, in addition to the different bright-end slope, the composite LF for cluster members also has a different shape with respect to the one for the field.

\section{Summary and Conclusions} \label{conclusions}

In this work, we build an IR composite luminosity function for the apparent magnitude of cluster members using WISE data. We identify the cluster members by matching the position of the WISE sources with the cluster members in the redMaPPer DR8 SDSS catalog. We consider $9$ redshift bins containing $100$ clusters each, in two different samples. The reference sample contains clusters with $\lambda > 45$, while the control sample contains clusters with $45 < \lambda < 90$. For the two samples, we consider both $w1$ and $w2$ apparent magnitudes. We fit the normalized composite LF with a double power law and provide best-fit parameters for the transition magnitude $M^{*}$ and the bright and faint-end slopes $\beta$ and $\alpha$, respectively. From our analysis we can draw the following conclusions:

\begin{enumerate}
\item The composite LF for cluster members in WISE is characterized by a value of the bright-end slope $\beta$ which is significantly steeper than the one for the field. For clusters in the reference sample, $\beta$ in the $w1$ band ranges from $3.49$ to $3.76$ at different redshifts, and it is compatible with a constant value of $3.54$. For clusters in the control sample, $\beta$ shows a mild negative trend with increasing redshift, with values ranging from $3.17$ to $3.60$. On the other end, the field is described by a slope between $2.8$ and $3.1$, which is significantly different from the one of clusters, independently on the richness of the sample. This is a clear signature that can be used to identify cluster members in WISE data.
 \item The faint end slope $\alpha$ of the members composite LF has a value between $2.0$ and $3.0$, remarkably different from $\beta$. This is an indication of the existence of a transition between the bright end and the faint end that is not present in the field, where the value for $\alpha$ is similar to $\beta$. The trend with redshift is less clear than in the bright end, but the existence of this transition itself can be used as a secondary characterization when searching for clusters in the WISE data. 
\item The transition apparent magnitude $M^{*}$ is increasing (shifting towards fainter luminosities) with increasing redshift. This means that, even if the global shape of the composite LF (and in particular the bright end) is the same at all redshifts, there is a shift in the transition magnitude. This is not just a redshift effect, as we discuss in Appendix \ref{app_absolute}. This fact can be used as an a posteriori additional information to constrain the redshift of clusters detected with WISE data.
\end{enumerate}
      

With the composite LF discussed in this work, we provide an ingredient for detecting clusters in the WISE survey with matched-filter algorithms that will be implemented in subsequent studies.

\begin{acknowledgements}
The authors thank Alexandre Beelen, Herv\'e Dole and Guillaume Hurier for fruitful discussion.
C.D.B. also thanks Carlotta Gruppioni, Elena Zucca and Micol Bolzonella for useful discussion on infrared astronomy.
This work was supported by a grant from R\'egion Ile-de-France DIM-ACAV. This research made use of the SZ-Cluster Database operated by the Integrated Data and Operation Center (IDOC) at the Institut d'Astrophysique Spatiale (IAS) under contract with CNES and CNRS. It also used data products from the Wide-field Infrared Survey Explorer, which is a joint project of the University of California, Los Angeles, and the Jet Propulsion Laboratory/California Institute of Technology, funded by the National Aeronautics and Space Administration.  This research made use of Astropy, a community-developed core Python package for Astronomy \citep{2013A&A...558A..33A}, and Matplotlib library \citep{Hunter:2007}.

\end{acknowledgements}


\bibliographystyle{aa} 
\bibliography{cristiano} 

\begin{thebibliography}{43}
\expandafter\ifx\csname natexlab\endcsname\relax\def\natexlab#1{#1}\fi
\expandafter\ifx\csname url\endcsname\relax
  \def\url#1{{\tt #1}}\fi
\expandafter\ifx\csname urlprefix\endcsname\relax\def\urlprefix{URL }\fi

\bibitem[{{Alam} et~al.(2015){Alam}, {Albareti}, {Allende Prieto}
  et~al.}]{2015ApJS..219...12A}
{Alam} S., {Albareti} F.D., {Allende Prieto} C., et~al., Jul. 2015, \apjs, 219,
  12

\bibitem[{{Ascaso} et~al.(2012){Ascaso}, {Wittman}, \&
  {Ben{\'{\i}}tez}}]{2012MNRAS.420.1167A}
{Ascaso} B., {Wittman} D., {Ben{\'{\i}}tez} N., Feb. 2012, \mnras, 420, 1167

\bibitem[{{Astropy Collaboration} et~al.(2013){Astropy Collaboration},
  {Robitaille}, {Tollerud} et~al.}]{2013A&A...558A..33A}
{Astropy Collaboration}, {Robitaille} T.P., {Tollerud} E.J., et~al., Oct. 2013,
  \aap, 558, A33

\bibitem[{{Babbedge} et~al.(2006){Babbedge}, {Rowan-Robinson}, {Vaccari}
  et~al.}]{2006MNRAS.370.1159B}
{Babbedge} T.S.R., {Rowan-Robinson} M., {Vaccari} M., et~al., Aug. 2006,
  \mnras, 370, 1159

\bibitem[{{Bell} \& {de Jong}(2001)}]{2001ApJ...550..212B}
{Bell} E.F., {de Jong} R.S., Mar. 2001, \apj, 550, 212

\bibitem[{{Bellagamba} et~al.(2017){Bellagamba}, {Roncarelli}, {Maturi}, \&
  {Moscardini}}]{2017arXiv170503029B}
{Bellagamba} F., {Roncarelli} M., {Maturi} M., {Moscardini} L., May 2017, ArXiv
  e-prints

\bibitem[{{Bleem} et~al.(2015){Bleem}, {Stalder}, {de Haan}
  et~al.}]{2015ApJS..216...27B}
{Bleem} L.E., {Stalder} B., {de Haan} T., et~al., Feb. 2015, \apjs, 216, 27

\bibitem[{{B{\"o}hringer} et~al.(2004){B{\"o}hringer}, {Schuecker}, {Guzzo}
  et~al.}]{2004A&A...425..367B}
{B{\"o}hringer} H., {Schuecker} P., {Guzzo} L., et~al., Oct. 2004, \aap, 425,
  367

\bibitem[{{Burenin}(2015)}]{2015AstL...41..167B}
{Burenin} R.A., May 2015, Astronomy Letters, 41, 167

\bibitem[{{Burenin}(2017)}]{2017AstL...43..507B}
{Burenin} R.A., Aug. 2017, Astronomy Letters, 43, 507

\bibitem[{{Cavaliere} \& {Fusco-Femiano}(1976)}]{1976A&A....49..137C}
{Cavaliere} A., {Fusco-Femiano} R., May 1976, \aap, 49, 137

\bibitem[{{Colless}(1989)}]{1989MNRAS.237..799C}
{Colless} M., Apr. 1989, \mnras, 237, 799

\bibitem[{{de Filippis} et~al.(2011){de Filippis}, {Paolillo}, {Longo}
  et~al.}]{2011MNRAS.414.2771D}
{de Filippis} E., {Paolillo} M., {Longo} G., et~al., Jul. 2011, \mnras, 414,
  2771

\bibitem[{{De Propris} et~al.(2003){De Propris}, {Colless}, {Driver}
  et~al.}]{2003MNRAS.342..725D}
{De Propris} R., {Colless} M., {Driver} S.P., et~al., Jul. 2003, \mnras, 342,
  725

\bibitem[{{Gavazzi} et~al.(1996){Gavazzi}, {Pierini}, \&
  {Boselli}}]{1996A&A...312..397G}
{Gavazzi} G., {Pierini} D., {Boselli} A., Aug. 1996, \aap, 312, 397

\bibitem[{{Goto} et~al.(2010){Goto}, {Takagi}, {Matsuhara}
  et~al.}]{2010A&A...514A...6G}
{Goto} T., {Takagi} T., {Matsuhara} H., et~al., May 2010, \aap, 514, A6

\bibitem[{{Hansen} et~al.(2005){Hansen}, {McKay}, {Wechsler}
  et~al.}]{2005ApJ...633..122H}
{Hansen} S.M., {McKay} T.A., {Wechsler} R.H., et~al., Nov. 2005, \apj, 633, 122

\bibitem[{{Hasselfield} et~al.(2013){Hasselfield}, {Hilton}, {Marriage}
  et~al.}]{2013JCAP...07..008H}
{Hasselfield} M., {Hilton} M., {Marriage} T.A., et~al., Jul. 2013, \jcap, 7,
  008

\bibitem[{Hunter(2007)}]{Hunter:2007}
Hunter J.D., 2007, Computing In Science \& Engineering, 9, 90

\bibitem[{{Kettlety} et~al.(2018){Kettlety}, {Hesling}, {Phillipps}
  et~al.}]{2018MNRAS.473..776K}
{Kettlety} T., {Hesling} J., {Phillipps} S., et~al., Jan. 2018, \mnras, 473,
  776

\bibitem[{{Kurcz} et~al.(2016){Kurcz}, {Bilicki}, {Solarz}
  et~al.}]{2016A&A...592A..25K}
{Kurcz} A., {Bilicki} M., {Solarz} A., et~al., Jul. 2016, \aap, 592, A25

\bibitem[{{Lake} et~al.(2017{\natexlab{a}}){Lake}, {Wright}, {Assef}
  et~al.}]{2017arXiv170207829L}
{Lake} S.E., {Wright} E.L., {Assef} R.J., et~al., Feb. 2017{\natexlab{a}},
  ArXiv e-prints

\bibitem[{{Lake} et~al.(2017{\natexlab{b}}){Lake}, {Wright}, {Tsai}, \&
  {Lam}}]{2017AJ....153..189L}
{Lake} S.E., {Wright} E.L., {Tsai} C.W., {Lam} A., Apr. 2017{\natexlab{b}},
  \aj, 153, 189

\bibitem[{{Lan} et~al.(2016){Lan}, {M{\'e}nard}, \& {Mo}}]{2016MNRAS.459.3998L}
{Lan} T.W., {M{\'e}nard} B., {Mo} H., Jul. 2016, \mnras, 459, 3998

\bibitem[{{Le Floc'h} et~al.(2005){Le Floc'h}, {Papovich}, {Dole}
  et~al.}]{2005ApJ...632..169L}
{Le Floc'h} E., {Papovich} C., {Dole} H., et~al., Oct. 2005, \apj, 632, 169

\bibitem[{{Licitra} et~al.(2017){Licitra}, {Mei}, {Raichoor}, {Erben}, \&
  {Hildebrandt}}]{2017arXiv170504331L}
{Licitra} R., {Mei} S., {Raichoor} A., {Erben} T., {Hildebrandt} H., May 2017,
  ArXiv e-prints

\bibitem[{{Moretti} et~al.(2015){Moretti}, {Bettoni}, {Poggianti}
  et~al.}]{2015A&A...581A..11M}
{Moretti} A., {Bettoni} D., {Poggianti} B.M., et~al., Sep. 2015, \aap, 581, A11

\bibitem[{{Murata} et~al.(2017){Murata}, {Nishimichi}, {Takada}
  et~al.}]{2017arXiv170701907M}
{Murata} R., {Nishimichi} T., {Takada} M., et~al., Jul. 2017, ArXiv e-prints

\bibitem[{{Navarro} et~al.(1996){Navarro}, {Frenk}, \&
  {White}}]{1996ApJ...462..563N}
{Navarro} J.F., {Frenk} C.S., {White} S.D.M., May 1996, \apj, 462, 563

\bibitem[{{Planck Collaboration} et~al.(2011){Planck Collaboration}, {Ade},
  {Aghanim} et~al.}]{2011A&A...536A...8P}
{Planck Collaboration}, {Ade} P.A.R., {Aghanim} N., et~al., Dec. 2011, \aap,
  536, A8

\bibitem[{{Planck Collaboration} et~al.(2016{\natexlab{a}}){Planck
  Collaboration}, {Ade}, {Aghanim} et~al.}]{2016A&A...594A..24P}
{Planck Collaboration}, {Ade} P.A.R., {Aghanim} N., et~al., Sep.
  2016{\natexlab{a}}, \aap, 594, A24

\bibitem[{{Planck Collaboration} et~al.(2016{\natexlab{b}}){Planck
  Collaboration}, {Ade}, {Aghanim} et~al.}]{2016A&A...594A..27P}
{Planck Collaboration}, {Ade} P.A.R., {Aghanim} N., et~al., Sep.
  2016{\natexlab{b}}, \aap, 594, A27

\bibitem[{{Popesso} et~al.(2006){Popesso}, {Biviano}, {B{\"o}hringer}, \&
  {Romaniello}}]{2006A&A...445...29P}
{Popesso} P., {Biviano} A., {B{\"o}hringer} H., {Romaniello} M., Jan. 2006,
  \aap, 445, 29

\bibitem[{{Rozo} et~al.(2015){Rozo}, {Rykoff}, {Becker}, {Reddick}, \&
  {Wechsler}}]{2015MNRAS.453...38R}
{Rozo} E., {Rykoff} E.S., {Becker} M., {Reddick} R.M., {Wechsler} R.H., Oct.
  2015, \mnras, 453, 38

\bibitem[{{Rykoff} et~al.(2014){Rykoff}, {Rozo}, {Busha}
  et~al.}]{2014ApJ...785..104R}
{Rykoff} E.S., {Rozo} E., {Busha} M.T., et~al., Apr. 2014, \apj, 785, 104

\bibitem[{{Rykoff} et~al.(2016){Rykoff}, {Rozo}, {Hollowood}
  et~al.}]{2016ApJS..224....1R}
{Rykoff} E.S., {Rozo} E., {Hollowood} D., et~al., May 2016, \apjs, 224, 1

\bibitem[{{Salvati} et~al.(2017){Salvati}, {Douspis}, \&
  {Aghanim}}]{2017arXiv170800697S}
{Salvati} L., {Douspis} M., {Aghanim} N., Aug. 2017, ArXiv e-prints

\bibitem[{{Schechter}(1976)}]{1976ApJ...203..297S}
{Schechter} P., Jan. 1976, \apj, 203, 297

\bibitem[{{Simet} et~al.(2017){Simet}, {McClintock}, {Mandelbaum}
  et~al.}]{2017MNRAS.466.3103S}
{Simet} M., {McClintock} T., {Mandelbaum} R., et~al., Apr. 2017, \mnras, 466,
  3103

\bibitem[{{Stanford} et~al.(2014){Stanford}, {Gonzalez}, {Brodwin}
  et~al.}]{2014ApJS..213...25S}
{Stanford} S.A., {Gonzalez} A.H., {Brodwin} M., et~al., Aug. 2014, \apjs, 213,
  25

\bibitem[{{Sunyaev} \& {Zeldovich}(1972)}]{1972CoASP...4..173S}
{Sunyaev} R.A., {Zeldovich} Y.B., Nov. 1972, Comments on Astrophysics and Space
  Physics, 4, 173

\bibitem[{{Wright} et~al.(2010){Wright}, {Eisenhardt}, {Mainzer}
  et~al.}]{2010AJ....140.1868W}
{Wright} E.L., {Eisenhardt} P.R.M., {Mainzer} A.K., et~al., Dec. 2010, \aj,
  140, 1868

\bibitem[{{Zucca} et~al.(2009){Zucca}, {Bardelli}, {Bolzonella}
  et~al.}]{2009A&A...508.1217Z}
{Zucca} E., {Bardelli} S., {Bolzonella} M., et~al., Dec. 2009, \aap, 508, 1217

\end{thebibliography}


\begin{appendix} 
\section{Absolute magnitude composite LF} \label{app_absolute}

For the $w1$ band we also build the absolute magnitude composite LF by using the cluster redshifts provided by the redMaPPer SDSS catalog. For every $i$-th galaxy with apparent magnitude $w_i$, we evaluate the absolute magnitude $W_i$ as:

\begin{equation}
W_i = w_i - DM(z_i) - K(z_i) \ ,
\end{equation}

\noindent where $z_i$ is the cluster redshift,

\begin{equation}
DM(z_i) = 5 \log \left( \frac{1+z_i}{10 {\rm \ pc}}  \int_{0}^{z_i}  \frac{c}{H_0 \sqrt{\Omega_m (1+z)^3 + \Omega_{\Lambda}}} dz \right) \ 
\end{equation}

\noindent is the distance modulus and

\begin{equation}
K(z_i) = -2.5 \log (1+z_i) 
\end{equation}

\noindent is the K-correction for WISE (Huang, \url{https://zenodo.org/record/31255/files/CarolineHuang.pdf}). We then proceed with the fit as for the apparent magnitude composite LF. For the normalization, we use $W1_{norm} = -24$. We report the results in Table \ref{w1_abs_24_45-500}.

\begin{table*}[h!]
\begin{center}
\caption{Best-fit parameters for the double power law, equation (\ref{double_power_law_m}), fit to the normalized absolute $w1$ composite LF for WISE data (with $W1_{norm} = -24$) for our reference sample ($\lambda>45$). \label{w1_abs_24_45-500}}
\begin{tabular}{cccccccccc}
\hline
\hline
$z$ & $\phi^{*}$ & $\sigma_{\phi^{*}}$ & $\alpha$ & $\sigma_{\alpha}$ & $\beta$ & $\sigma_{\beta}$ & $M^{*}$ & $\sigma_{M^{*}}$ & $\chi^2_{red}$ \\
\hline
$0.10-0.15$ & $11.71$ & $0.96$ & $2.61$ & $0.09$ & $3.63$ & $0.09$ & $-24.52$ & $0.07$ & $0.487$ \\
\hline
$0.15-0.20$ & $14.01$ & $0.85$ & $2.25$ & $0.08$ & $3.61$ & $0.07$ & $-24.60$ & $0.05$ & $$0.80 \\ 
\hline
$0.20-0.25$ & $13.89$ & $0.69$ & $2.36$ & $0.09$ & $3.49$ & $0.09$ & $-24.84$ & $0.07$ & $0.97$ \\ 
\hline
$0.25-0.30$ & $8.14$ & $1.15$ & $2.95$ & $0.10$ & $3.78$ & $0.15$ & $-25.49$ & $0.12$ & $1.70$ \\
\hline
$0.30-0.35$ & $11.67$ & $0.54$ & $2.21$ & $0.09$ & $3.54$ & $0.11$ & $-25.36$ & $0.07$ & $0.61$ \\ 
\hline
$0.35-0.40$ & $8.40$ & $--$ & $2.91$ & $--$ & $3.69$ & $0.09$ & $-25.75$ & $--$ & $2.20$ \\ 
\hline
$0.40-0.45$ & $8.87$ & $--$ & $2.53$ & $--$ & $3.64$ & $0.09$ & $-27.75$ & $--$ & $1.06$ \\ 
\hline
$0.45-0.50$ & $8.46$ & $0.50$ & $2.13$ & $0.11$ & $3.73$ & $0.16$ & $-25.95$ & $0.07$ & $1.47$ \\ 
\hline
$0.50-0.55$ & $5.72$ & $--$ & $2.94$ & $--$ & $3.57$ & $0.14$ & $-26.25$ & $--$ & $6.44$ \\
\hline
\hline
\end{tabular}
\end{center}
\end{table*}

\begin{figure}
\centering
\hbox{
 \includegraphics[width=0.45\textwidth]{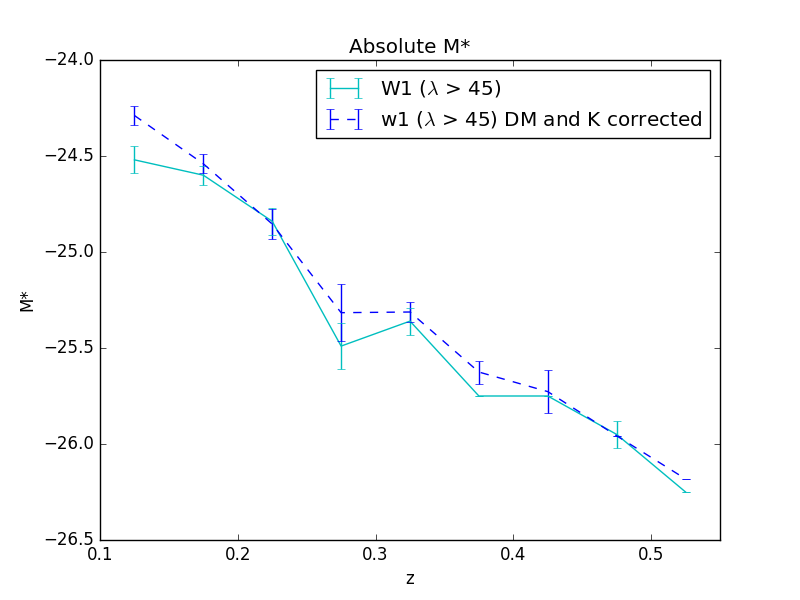}
}
\hbox{
}
\caption{Evolution with redshift of the best-fit absolute $M^{*}$ for $W1$ ($\lambda>45$) [cyan full-line] and $w1$ ($\lambda>45$) corrected for distance modulus and K-correction [blue dashed-line]. Error bars mark one standard deviation.}
\label{w_abs_star}
\end{figure}

In Figure \ref{w_abs_star} we compare the best-fit absolute $M^{*}$ with the best-fit apparent $M^{*}$ corrected for distance modulus and K-correction. There are small differences due to the assignation of few members to different magnitude bins, but the trend is well reproduced, as expected.
For the absolute $M^{*}$ the trend with redshift that we found for the apparent $M^{*}$ does not disappear, indicating that what we see in Figure \ref{w_app_star} is not just a redshift effect. Actually, the transition happens at a brighter absolute $M^{*}$ for high-redshift clusters than for low-redshift ones. This means that the composite LF is not completely self-similar at different redshifts, despite the results we discussed for the bright-end and faint-end slopes remain valid when considering absolute magnitudes.

\end{appendix}

\end{document}